\documentclass[runningheads, a4paper]{llncs}
\usepackage[T1]{fontenc}
\usepackage{graphicx}
\usepackage{siunitx}
\usepackage{color}
\usepackage{hyperref}

\providecommand{\CaptionTitle}[1]{\textbf{#1}}
\providecommand{\IfDocumentMetadataTF}[2]{#2}
\IfDocumentMetadataTF{}{\usepackage{accsupp}}
\makeatletter
\IfDocumentMetadataTF{}{%
  \newcommand*{\PublicationEndAltText}{\EndAccSupp{}}%
  \define@key{Gin}{alt}{%
    \BeginAccSupp{method=pdfstringdef,unicode,Alt={#1}}%
    \aftergroup\PublicationEndAltText
  }%
}
\makeatother

\makeatletter
\@ifundefined{xmpquote}{}{}
\makeatother
\hypersetup{
  pdftitle={WAMpy: Efficient Synthesis of Prolog Programs in Python},
  pdfauthor={Dominik Magiera; Lukas Röhrig; Frank Jäkel},
  pdflang={en-US},
  hidelinks,
  pdfcreator={LaTeX via pandoc}}

\begin{document}
\title{WAMpy: Efficient Synthesis of Prolog Programs in Python}
\author{%
Dominik Magiera\orcidID{0009-0002-0643-9973}%
 \and
Lukas Röhrig\orcidID{0009-0009-9664-8226}%
 \and
Frank Jäkel\orcidID{0000-0002-1355-7663}%
}
\authorrunning{D. Magiera et al.}
\institute{%
Centre for Cognitive Science \& Institute of Psychology, Technische Universität Darmstadt, Germany%
}
\maketitle              %

\begin{abstract}
We present WAMpy, a Python framework optimized for synthesizing Prolog programs. Unlike general-purpose Prolog systems,
WAMpy targets workloads that repeatedly generate and evaluate small candidate programs. WAMpy compiles Prolog clauses
into \emph{NumPy} array-based \emph{WAM} instructions and supports partial recompilation of hypotheses against fixed
background knowledge. Performance-critical routines are accelerated using \emph{Numba} just-in-time (JIT) compilation.
In a benchmark of repeated compilation-and-evaluation workloads, WAMpy improves end-to-end performance compared with
SWI-Prolog accessed from Python using Janus.

\emph{WAMpy is available on \href{https://github.com/cognitive-modeling/WAMpy}{github.com/cognitive-modeling/WAMpy}.}

\end{abstract}

\section{Usage and Implementation}\label{usage-and-implementation}

Program synthesis constructs programs from incomplete specifications, such as examples or constraints
\cite{Flener_1995}. This process often involves repeatedly compiling and evaluating candidate programs while the
background knowledge remains unchanged. Existing general-purpose Prolog systems such as \emph{SWI-Prolog}
\cite{Wielemaker_2003} are not optimized for this workload. We therefore present \emph{WAMpy}, a specialized execution
framework based on the \emph{Warren Abstract Machine (WAM)} \cite{Warren_1983,Ait-Kaci_1991}. WAMpy is implemented in
Python and uses Prolog as its source language, enabling direct integration with program-synthesis and machine-learning
workflows. It compiles programs into a targeted subset of array-based WAM instructions. Performance-critical routines
are accelerated using Numba JIT compilation \cite{Lam_etal_2015}.

\subsubsection{Usage.}\label{usage.}

WAMpy is used as a Python library rather than as a standalone Prolog interpreter. The following example compiles Prolog
source into WAM instructions and evaluates a query through the high-level API: \[
\vbox{
\hbox{\verb|import wampy as wam|}
\hbox{\verb|prolog = wam.Prolog("parent(anakin, luke).")|}
\hbox{\verb|list(prolog.query("parent(X, luke)."))|}
}
\] The high-level API manages parsing, compilation, and query backtracking.

\subsubsection{Implementation.}\label{implementation.}

\begin{figure}
\centering
\includegraphics[width=4.75in,alt={Three-panel diagram showing WAMpy\textquotesingle s source-to-runtime pipeline,
NumPy-based term representation, and partial-recompilation process.}]{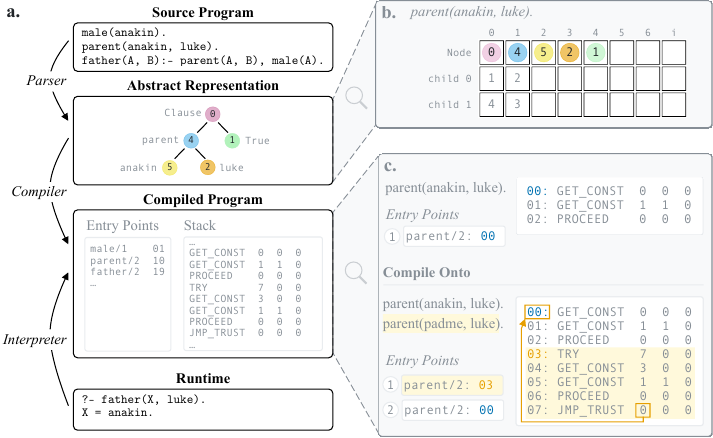}
\caption[WAMpy compilation and execution workflow.]{\CaptionTitle{WAMpy compilation and execution workflow.}
\textbf{a.}~Pipeline from Prolog source code to query results. \textbf{b.}~NumPy-based representation of a parsed term.
\textbf{c.}~Partial recompilation adds a candidate clause to precompiled background knowledge.}\label{fig:overview}
\end{figure}

Uncompiled programs are represented as fixed-size arrays of top-level term trees. Each parsed term is stored as a
separate AST with its own root node and integer identifier. Programs can be parsed from Prolog source code, but are
primarily intended to be modified directly at the AST level during synthesis. Each program is constrained by fixed upper
bounds on the number of terms, nodes per term, and children per node. Memory is allocated according to these bounds, so
all programs share the same array layout. This increases memory use, but enables efficient processing by Numba-compiled
routines. Internally, programs use fixed-size NumPy arrays: a symbol array indexed by term and node, and a child array
indexed by term, node, and child slot. The number of child slots is configured in advance
(Fig.~\hyperref[fig:overview]{\ref*{fig:overview}b}).

Compilation lowers these AST's into a two-dimensional array, where each row represents one instruction in our WAM
subset. Each instruction consists of an opcode followed by three arguments. The compiler also constructs a predicate
entry table that maps functor-arity pairs to program locations. The resulting instruction array and predicate entry
table are illustrated in Fig.~\hyperref[fig:overview]{\ref*{fig:overview}a}. Multiple clauses for the same predicate are
handled using WAM-style choice instructions to support backtracking.

At runtime, queries are evaluated using \emph{Selective Linear Definite} (SLD) clause resolution
\cite{Kowalski_Kuehner_1971}, where unification produces bindings and backtracking explores alternative clauses. The
framework provides built-in tracing of predicate calls during execution, which can be used to evaluate candidate
hypotheses.

\subsubsection{Efficient Partial Recompilation.}\label{efficient-partial-recompilation.}

WAMpy reduces compilation overhead by compiling background knowledge once and recompiling only candidate hypotheses.
Candidate clauses are appended to the fixed instruction sequence and tried before background clauses for the same
predicate. In the single-clause case shown in Fig.~\hyperref[fig:overview]{\ref*{fig:overview}c}, the compiler appends a
\texttt{JMP\_TRUST} bridge that transfers backtracking to the precompiled background clause. Predicates with multiple
background clauses use \texttt{JMP\_RETRY} to preserve further alternatives.

\subsubsection{Limitations.}\label{limitations.}

WAMpy implements only an X-register subset of the WAM \cite{Ait-Kaci_1991}, for repeated compilation and evaluation in
logic program-synthesis workloads. This subset supports atoms, variables, compound structures, predicate calls, tail
calls, unification, and linear clause backtracking. It does not provide a general SWI-Prolog built-in predicate library,
including facilities such as arithmetic and meta-calls.

\section{Evaluation}\label{evaluation}

We use pytest for automated validation of WAMpy's main components and end-to-end behavior. The tests run Prolog programs
to check that parsing, compilation, unification, backtracking, and query evaluation produce the expected answers.

\subsubsection{Benchmark.}\label{benchmark.}

\begin{figure}
\centering
\includegraphics[width=4.75in,alt={Two-panel benchmark comparing the end-to-end runtime of WAMpy and Janus SWI-Prolog
and the compilation time of full and partial WAMpy compilation.}]{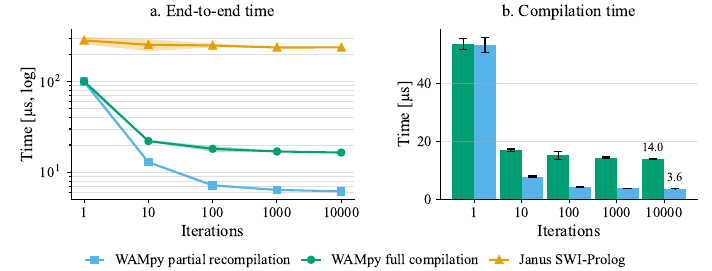}
\caption[Performance benchmark of WAMpy and Janus SWI-Prolog.]{\CaptionTitle{Performance benchmark of WAMpy and Janus
SWI-Prolog.} \textbf{a.}~End-to-end runtime, including compilation and query execution. \textbf{b.}~WAMpy compilation
time for full-program compilation and partial hypothesis-only recompilation.}\label{fig:benchmark}
\end{figure}

We evaluate WAMpy on a small family-relation benchmark that mimics a synthesis loop. The program contains 13 gender
facts, 13 parent facts, and seven rules.\footnote{The full source is available online:
  \url{https://github.com/cognitive-modeling/WAMpy}.} For partial recompilation, the final three rules are treated as
the hypothesis, while the remaining 30 clauses form the background knowledge. Evaluation is simulated by repeatedly
querying \texttt{father(X,\ luke)}. We compare against Janus 1.5.3\footnote{We use \texttt{janus-swi} 1.5.3 with
  \texttt{SWI-Prolog} 10.0.2, available on \href{https://pypi.org/project/janus-swi}{PyPI}.}, using SWI-Prolog 10.0.2
and Python 3.13, as a Python-hosted Prolog baseline rather than evaluating standalone SWI-Prolog. All benchmarks were
performed on a MacBook Pro with an Apple M3 Max processor (16 CPU cores) and 64 gigabytes of unified memory. For each
iteration count \(N\), the program was compiled and evaluated \(N\) times, with 10 samples collected per setting. Both
WAMpy compilation modes outperform Janus in end-to-end runtime, including compilation and query execution, as shown in
Fig.~\hyperref[fig:benchmark]{\ref*{fig:benchmark}a}. With increasing iterations, compilation costs become amortized,
leaving Janus nearly constant at around \qty{240}{\micro\second} while WAMpy decreases to
\qty{16.51 \pm 0.13}{\micro\second} with full compilation and \qty{6.17 \pm 0.03}{\micro\second} with partial
recompilation. Fig.~\hyperref[fig:benchmark]{\ref*{fig:benchmark}b} isolates WAMpy's compilation time. Both compilation
modes start at around \qty{54}{\micro\second}, but partial recompilation increasingly outperforms full compilation,
reaching \qty{3.64 \pm 0.03}{\micro\second} compared with \qty{13.98 \pm 0.13}{\micro\second}.

\subsubsection{Conclusion.}\label{conclusion.}

WAMpy is a modern Python framework for program synthesis built on NumPy data structures and Numba just-in-time (JIT)
compilation. For program synthesis workloads, WAMpy is approximately 38.5 times faster than Janus SWI-Prolog by
leveraging partial recompilation. Future work will extend Prolog support, explore GPU acceleration with Numba CUDA, and
evaluate larger inductive logic programming benchmarks.

\begin{credits}
\subsubsection{\ackname}
This work was supported by the Honda Research Institute Europe (HRI-EU).

\subsubsection{\discintname}
The authors have no competing interests to declare that are relevant to the content of this article.

\end{credits}
\bibliographystyle{splncs04_unsrt}
\bibliography{references}
\end{document}